# Neutral genomic regions refine models of recent rapid human population growth

Elodie Gazave[1,6], Li Ma[1,6,#], Diana Chang[1,6], Alex Coventry[2], Feng Gao[1], Donna Muzny[3], Eric Boerwinkle[3,4], Richard A. Gibbs[3], Charles F. Sing[5], Andrew G. Clark[2,1], Alon Keinan[1,*]

[1] Department of Biological Statistics and Computational Biology, Cornell University, Ithaca, NY 14853, USA

[2] Department of Molecular Biology and Genetics, Cornell University, Ithaca, NY 14853, USA.

[3] Department of Molecular and Human Genetics, Baylor College of Medicine, Houston, TX 77030, USA

[4] Human Genetics Center, Health Science Center, University of Texas, Houston, TX 77030, USA

[5] Department of Human Genetics, University of Michigan, Ann Arbor, MI 48105, USA

[6] These authors contributed equally to this work

[#] Current Address: Department of Animal and Avian Sciences, University of Maryland, College Park, MD 20742, USA

[*] To whom correspondence should be addressed. E-mail: ak735@cornell.edu

**Corresponding author**: Alon Keinan, 102C Weill Hall, Cornell University, Ithaca NY 14853-7202, USA; Phone: 607-254-1328; ak735@cornell.edu

# Abstract

Human populations have experienced dramatic growth since the Neolithic revolution. Recent studies that sequenced a very large number of individuals observed an extreme excess of rare variants, and provided clear evidence of recent rapid growth in effective population size, though estimates have varied greatly among studies. All these studies were based on protein-coding genes, in which variants are also impacted by natural selection. In this study, we introduce targeted sequencing data for studying recent human history with minimal confounding by natural selection. We sequenced loci very far from genes that meet a wide array of additional criteria such that mutations in these loci are putatively neutral. As population structure also skews allele frequencies, we sequenced a sample of relatively homogeneous ancestry by first analyzing the population structure of 9,716 European Americans. We employed very high coverage sequencing to reliably call rare variants, and fit an extensive array of models of recent European demographic history to the site frequency spectrum. The best-fit model estimates ~3.4% growth per generation during the last ~140 generations, resulting in a population size increase of two orders of magnitude. This model fits the data very well, largely due to our observation that assumptions of more ancient demography can impact estimates of recent growth. This observation and results also shed light on the discrepancy in demographic estimates among recent studies.

# Significance Statement

The recent rapid growth of human populations predicts that a large number of genetic variants in populations today are very rare, i.e. appear in a small number of individuals. This effect is similar to that of purifying selection, which drives deleterious alleles to become rarer. Recent studies of the genetic signature left by rapid growth were confounded by purifying selection since they focused on genes. Here, to study recent human history with minimal confounding by selection, we sequenced and examined genetic variants very far from genes. These data point to the human population size growing by about 3.4% per generation over the last 3-4 thousand years, resulting in an over 100-fold increase in population size over that epoch.

# Introduction

Archeological and historical records reveal that modern human populations have experienced a dramatic growth, likely driven by the Neolithic revolution about 10,000 years ago (1, 2). Since then, the worldwide human population size has increased at a fast pace, and faster yet in the last ~2000 years, giving rise to today's population in excess of 7 billion people (3, 4). A central question in population genetics is how such demographic events affect the effective size ($N_e$) of populations over time, and as a consequence how they have shaped extant patterns of genetic variation [effective population size, which is typically smaller than the census size, determines the genetic properties of a population (5)]. Focusing often on human populations of European descent, estimates of $N_e$ from genetic variation have been traditionally on the order of 10,000 individuals (6-11), though higher and lower estimates have also been obtained (12-16). More recent studies based on resequencing data from a relatively small number of individuals have also considered recent population growth in fitting models to the observed site frequency spectrum (SFS), and reported as much as 0.5% increase in $N_e$ per generation, culminating in a $N_e$ of a few tens of thousands today (13, 14). It has been recently hypothesized that these studies could not capture the full scope of population growth since a larger sample size of individuals is needed to observe SNVs (single nucleotide variants) that arose during the recent epoch of growth (4).

With extreme recent population growth as experienced by human populations, the vast majority of SNVs are expected to be very young and rare, i.e. of very low allele frequency (4). Indeed, several recent sequencing studies with very large numbers of individuals have observed an unprecedented excess in the proportion of rare SNVs (17-19). Fitting models to the SFS, these studies have captured a clearer, more rapid recent population growth than earlier studies (17-19). At the same time, demographic estimates varied by as much as an order of magnitude between these studies (SI Appendix, Table S1).

Not all rare SNVs are as recent as others and a variant's selective effect plays an important part in its frequency. For instance, SNVs that are deleterious are on average younger than neutral variants of the same frequency, which was first posited by Maruyama (20, 21) and recently shown to be the case for humans (22). Purifying (negative) selection acting on deleterious alleles is expected to give rise to an excess of rare variants, which has been demonstrated for human populations (17, 19, 23). Thus, the genetic signature left by purifying selection on the SFS confounds the signature left by recent growth (24). To minimize this confounding effect, recent studies based on protein-coding genes considered for modeling solely synonymous SNVs, which do not modify the amino acid sequence (17-19). However, synonymous mutations have been shown to be targeted by natural selection, e.g. due to their impact on translation efficiency, translation accuracy, splicing, and folding energy (17, 19, 25-29). Hence, to study recent human genetic history with minimal effects of selection, it is not only desirable to consider accurate sequencing data from a large number of individuals, but also to focus on genomic regions in which mutations are putatively neutral, i.e. not affected by natural selection. Another potential confounder of demographic inference in recent studies has been population structure. Since both large-scale and fine-scale population structure exists in European populations (30-33), pooling individuals of different European ancestries can lead—when not accounted for in modeling (17-19)--to biases in the observed SFS and, consequently, in estimates of recent history (SI Appendix).

In this study, we aim to capture recent demographic history and estimate the magnitude of the recent growth experienced by humans, while limiting the confounding by natural selection and population structure. For this purpose, we selected a small set of genomic regions that are putatively neutral based on a wide array of criteria, the SFS of mutations in these regions is likely to reflect historical changes in $N_e$ rather than selection. We sequenced these regions in a large sample of individuals that share a relatively similar genetic ancestry within Europe. Very deep sequencing coverage allowed us to reliably observe even singletons (SNVs with an allele appearing in a single copy in the sample). Based on this dataset, we explored different models of recent European demographic history. Our best-fit model estimates a growth of 3.4% per generation over the last 141 generations, which is more rapid than estimated in recent large-scale studies (17, 19). We show that differences among previous models (17, 19)—and between these models and ours—can be partially explained by *a priori* assumptions about more ancient demographic events.

# Results

To estimate recent European demographic history with minimal confounding of natural selection, we sequenced genomic regions that we selected such that mutations therein would be as neutral as possible. Thus, the neutral regions (NR) dataset consists of loci that are at least 100,000 base-pairs (bp) and at least 0.1 centiMorgans (cM) away from coding or potentially coding sequences. The loci are also free of known copy number variants, segmental duplications, and loci that have been shown to have been under recent positive selection, and have minimal amounts of conserved and repetitive elements (Materials and Methods, SI Appendix). We sequenced a final set of 15 loci that met all these criteria, spanning a total of 216,240 bp (SI Appendix, Table S2).

In modeling a population's recent demographic history, if the sample of individuals trace their recent ancestry to different populations, even different European populations, the number of singletons and other rare variants can be biased (SI Appendix, Fig. S1). To sequence the neutral regions in a sample that minimizes such population structure, we applied principal component analysis (PCA) to whole-genome genotyping data of 9,716 individuals of European ancestry from the Atherosclerosis Risk in Communities (ARIC) cohort (34). This analysis revealed extensive population structure and we consequently sequenced 500 individuals that form a relatively homogenous cluster (Fig. 1a). Validating the choice of samples by comparing to a diverse panel of European populations from POPRES (35), these 500 individuals indeed show much less heterogeneity than the broader ARIC sample (Fig. 1b). They share a North-Western European ancestry that is similar to the POPRES UK sample (Fig. 1c).

Sequencing was carried out with Illumina HiSeq 100-bp paired-end for a median average coverage depth of 295X per individual after filtering of duplicated reads (SI Appendix, Fig. S2). We used UnifiedGenotyper (GATK) for variant detection and genotype calling (36, 37) and after applying strict filters to the raw calling, we obtained 1,834 high quality SNVs (SI Appendix). Of these variants, 62.5% have not been reported in dbSNP 135 (novel SNVs). The Ti/Tv (transition/transversion) ratio showed no indication of biases, for both all SNVs (Ti/Tv=2.22) and for novel SNVs alone (Ti/Tv = 2.29). None of the called SNVs presented significant departure from Hardy-Weinberg equilibrium. We further validated the quality of variant and genotype calling by comparing to those from whole-genome sequencing of the CHARGE-S project, which overlaps with a few individuals from our NR dataset (38). This validation supports the high quality of the NR dataset due to the very high sequencing coverage (SI Appendix). High-quality genotypes for at least 450 individuals were obtained for 95% of SNVs, which are used for presentation of the SFS throughout, while probabilistically subsampling 900 random chromosomes for each SNV (SI Appendix). We used the full set of SNVs for demographic modeling, where our approach accounts for missing data (Materials and Methods).

We compared the SFS from the NR dataset to that from the Exome Sequencing Project (ESP), which is of equivalent deep sequencing (SI Appendix, Fig. S3). We randomly subsampled the latter to 900 chromosomes and stratified by functional annotation (39). The more functional an annotation is considered, the more the ESP SFS deviates from that of the neutral regions: The highest agreement is observed for intronic and intergenic annotations from the ESP data, though still significantly different ($P < 10^{-3}$ for a goodness of fit test between the SFS of these SNVs and that of our NR data; SI Appendix, Table S3 and Fig. S4). Agreement is much worse between synonymous SNVs and the NR data ($P=5.4 \times 10^{-19}$), and worst for missense, splice, and nonsense SNVs ($P < 10^{-30}$ for each; SI Appendix, Fig. S4, Table S3). The lack of agreement between the SFS of ESP and NR is due to a higher proportion of very rare variants in the former (SI Appendix, Fig. S4), which is consistent with purifying selection playing a larger role in maintaining alleles at lower frequencies in and around genes. We further examined the NR SNVs in comparison to synonymous SNVs via GERP (Genomic Evolutionary Rate Profiling) scores (40), showing our data to be much more narrowly distributed around a score of 0, which corresponds to the absence of functional constraint (SI Appendix, Fig. S5). We note that though GERP scores may reflect the NR regions being selected to minimize conserved elements, applying the same conservation criteria to synonymous SNVs still results in a wider distribution of functional constraint (SI Appendix, Fig. S5). This set of results combined support our unique choice of regions and the relatively neutral nature of SNVs in these regions, as well as further validate our SNV and genotype calling pipeline.

Though the SFS of the NR dataset has a lower proportion of singletons compared to sites under purifying selection, it still exhibits a marked enrichment in the proportion of singletons (Fig. 2) compared to the expectation for a population that has remained at constant size throughout history (38.4% vs. 13.6%). This is consistent with the impact of recent population growth on the distribution of allele frequencies (4, 17-19). However, the SFS predicted by recently published demographic models of European populations that include recent exponential population growth (17, 19) do not closely match the SFS of the NR data (SI Appendix, Fig. S6). These models predict a higher proportion of rare variants and a smaller proportion of common variants than we observed, which is consistent with these models being based on synonymous SNVs with increased effect of purifying selection (17, 19). Hence, we next estimated the magnitude and duration of population growth by fitting the SFS of the NR data to several different models of recent history.

The first demographic model that we fit to the SFS consists of a recent exponential population growth with two free parameters: the time growth started and the extant $N_e$, with the growth assumed to continue into the present. More ancient demographic events, prior to the epoch of growth, including two population bottlenecks, were assumed to follow the model and estimates of (9) (SI Appendix, Table S1). The resulting model ('Model I') estimates the extant population size and, as a consequence, the growth rate, with very large uncertainty, as evident from the 95% confidence intervals (CI; Table 1), and the model does not fit the data very accurately (Fig. 2, Table 1).

Model I, similar to other recent models of population growth (17-19), assumed more ancient demography as fixed, with the intention of obtaining better resolution for estimating the two parameters of recent history. However, different recent models of population growth have assumed different models of ancient demography (SI Appendix, Table S1): Tennessen *et al.* (19) assumed that estimated previously by Gravel *et al.* (14), Nelson *et al.* (17) and Coventry *et al.* (18) that estimated by Schaffner *et al.* (15), and here that estimated by Keinan *et al.* (9). To test how sensitive Model I is to the details of assumed ancient history, we repeated fitting Model I while assuming each of the above ancient demographic models. Our results point to large differences among the 3 resulting models, with some parameters for recent demography being as much as an order of magnitude different (SI Appendix, Table S4). We stress that both data and methodology underlying these inferences are identical and, hence, conclude that the assumption of ancient demography has a major effect on estimating the timing and magnitude of recent population growth, which explains some of the differences among the recently published models (17-19).

To alleviate some of the sensitivity to ancient demography assumptions, we fit 'Model II' that extends Model I by adding an additional parameter for the effective population size just before exponential growth. This 3-parameter model fits the NR data significantly better than Model I ($P = 2.3 \times 10^{-6}$; Fig. 2). It estimates the ancestral $N_e$ before the growth to be 5,633 (CI of 4,400--7,100), markedly lower than the fixed value of 10,000 in Model I (Table 1). It estimates growth starting 141 (117--165) generations ago, which is a little earlier than in Model I, with a less rapid growth rate of 3.4% (2.4%--5.1%) per generation, which culminates in extant $N_e$ of 0.65 (0.3--2.87) million individuals (Table 1, Fig. 3-4).

To test whether this improved Model II can explain the differences between different assumed ancient demographic histories, we repeated fitting its 3 parameters to the NR data similarly to above with Model I. When ancient demography is assumed from Gravel *et al.* (14), this model fits the data significantly better (P = 2.9 × 10$^{-7}$) than the equivalent of Model I with the same ancient demography. Parameter estimates become practically identical to those of the above Model II based on Keinan *et al.* (9) (Fig. 4, SI Appendix, Tables S4-S5, Fig. S7). Model II based on Schaffner *et al.* (15) does not fit the data better than the respective Model I (P = 0.52) and provides the poorest fit of all three models (SI Appendix, Table S4-S5, Fig. S8). One notable difference in the model of Schaffner *et al.* (15) is that the timing of the second, European population bottleneck is assumed as fixed at over 2-fold that estimated in the other two models (9, 14) (SI Appendix, Table S1). The extant $N_e$ is almost identical for all these three models based on Model II, with best-fit estimates varying between 0.47 and 0.77 million individuals (Fig. 4, SI Appendix, Table S5). We conclude that by modeling the epoch prior to growth, the improved Model II is much less sensitive to assumptions about more ancient demography, and it goes a long way in closing the gap between different published models of recent growth (17, 19) and between these and Model I.

Archeological and historical records suggest that the growth of the human census population size has accelerated over time (3, 4). Our models thus far considered a single epoch of exponential growth, estimated to have started ~3500 years ago (assuming 25 years per generation). Hence, we considered several additional models in which growth can span two separate epochs with a different growth rate in each: 'Model III' with 3 parameters, 'Model IV' with 4 parameters, and all other possible 4-parameter models (Table 1). None of these more detailed models fit the data better than Model II (P > 0.86; Table 1, Fig. 2). They all estimate the second, more recent epoch of growth to be practically identical to the one estimated in Model II and the earlier epoch of growth to be equivalent to an epoch of constant population size (Table 1, SI Appendix, Fig. S7).

Finally, we investigated whether low statistical power due to a limited amount of data could explain the lack of an earlier epoch of growth. We simulated a scenario that is identical to Model II, except for the addition of an earlier epoch of milder growth and estimated how often Model III provides a significantly better fit than Model II (SI Appendix). We repeated this procedure for a variety of growth rates during the earlier epoch. Our modeling had a non-negligible statistical power of capturing two separate epochs of growth in all cases (SI Appendix, Fig. S9). Power is 86% when earlier growth rate is about half that of the more recent epoch. It decreases for milder growth during the first epoch—being as low as 25% for a growth rate of only 0.3% per generation—as well as when growth during the earlier epoch becomes similar to the recent rate (SI Appendix, Fig. S9). Overall, power is >60% for detecting two distinct epochs of growth as long as the growth rate during the earlier epoch is in the range of 0.6--1.8% (compared to 3.4% in the recent epoch), which is consistent with archaeological data (SI Appendix).

## Discussion

Recent studies have provided clear evidence that human populations have experienced recent explosive population growth, although detailed estimates of growth varied greatly (17-19). While all studies were in the context of medical genetic studies, hence based on the sequencing of protein-coding genes, here we generated a dataset with the sole purpose of accurately capturing rare putatively neutral variants for studying recent human history and population growth. As such, our NR data consists of several characteristics. First, since both demography and natural selection shape the distribution of allele frequencies, loci for the NR data were carefully chosen to minimize the influence of natural selection. Second, since population structure also affects the distribution of allele frequencies, the data consists of individuals with a homogenous European ancestry, similar to that of individuals from the UK. Third, since the genetic signature of growth is in rare variants, the data consists of a relatively large sample of 500 individuals. While some recent studies have considered a larger sample size, this was at the cost of studying a medical cohort with more heterogeneous ancestry. Fourth, to deal with the relatively high error rate of next-generation sequencing, we sequenced the neutral loci in all individuals to a very high coverage, which allows strict filtering and yields a set of very high quality SNVs. These characteristics combine to make the NR dataset ideally suited for population genetic studies of rare variants and recent history.

We used the NR data to consider an array of models of recent human demographic history, while showing that our results are consistent with being less confounded by natural selection. The best-fit model points to Europeans having experienced recent growth from an effective population size of about 4-7 thousand individuals as recently as 120--160 generations (3000--4000 years) ago. Growth over the last 3000-4000 years is estimated at an average rate of about 2--5% per generation, resulting in an overall increase in effective population size of two orders of magnitude. This model fits the data very well, but only after the realization that assumptions of ancient demography impact the estimate of recent population growth. We hypothesize that this is the case since previous models of ancient demographic history resulted in parameters that confound more recent and more ancient history (41), with the recent growth indirectly affecting them in a manner dependent upon sample size. This realization leads to the model we report here fitting much better than previous models of recent growth, and it sheds light on the discrepancies among the latter.

Motivated by archeological evidence of growth starting with the Neolithic revolution ~10,000 years ago and accelerating in the Common Era, we considered models that allow for acceleration of the rate of growth, but none supported such acceleration. One recent model considered two separate epochs of exponential growth (19). However, the first captures a slow recovery from the Eurasian population bottleneck ~23,000 years ago, with a weak growth rate of 0.3% that leads to an $N_e$ of only 9,208. This is similar to the instantaneous recovery from the population bottleneck in other models (14). Thus, to date no recent acceleration in the rate of growth that is along the lines suggested by archeological evidence has been observed in genetic data. Power calculations showed that with our data size and modeling framework, our results do not support a milder growth before it accelerated with >60% certainty. One explanation of our modeling not capturing two separate epochs of growth, other than limited statistical power, is that *effective* population size increases extremely slowly with the *census* population size, at least initially. While several factors contribute to these phenomenon, the particular increase in census population size with the Neolithic revolution has been accompanied by changing social structure that has led to increased variability in reproductive success; the advent of agriculture led to differential accumulation of richness, more notably in males, resulting in differential access to females compared to a hunter-gatherer life style (42). Increased variance in reproductive success results in relatively decreased effective population size. Perhaps jointly with other population processes associated with this social shift, e.g. changing generation time, either a lack of growth in effective population size initially or a milder one can be expected. This, in turn, can lead to our models having reduced power and thereby only capturing the more recent and more rapid growth.

In conclusion, we presented refined models of the recent explosive growth of a European population. These models can inform studies of natural selection (43-45), the architecture of complex diseases, and the methods that should best be used for genotype-phenotype mapping. We hope that our models and the public availability of our NR dataset will facilitate additional such studies. (Data available in dbSNP, with more detailed data at http://keinanlab.cb.bscb.cornell.edu.) However, models of recent demographic history are still limited to Europeans (17-19) and African Americans (19), and there is a need to extend them to additional populations. As the vast majority of rare variants are population-specific (27, 46, 47), such studies of additional populations will also facilitate better consideration of the replicability of genome-wide association studies results across populations.

# Materials and Methods

**Selection of individuals with shared European genetic ancestry**

Principal Component Analysis (PCA) was run on 9,716 European Americans from the ARIC cohort (34), using EIGENSOFT (48) on whole-genome genotyping data from the Affymetrix 6.0 genotyping array (dbGaP accession phs000090.v1.p1). Outliers of inferred non-European ancestry were removed, in addition to regions of extended linkage disequilibrium such as inversions (SI Appendix). A total of 500 individuals that were densely clustered together based on the first four principal components (PCs) were then chosen for sequencing. We tested for plate effect, which showed no correlation with the localization of the individuals on these PCs. We validated the ancestry of the 500 individuals by merging data from the Affymetrix 500k genotyping array of ten individuals from each POPRES population (35) with samples from ARIC and repeating PCA (SI Appendix).

**Choice of target putatively neutral regions**

To minimize the effect of selection, we considered genomic regions located at least 100,000 bp and 0.1 cM from any coding or potentially coding loci. We excluded genomic sequences containing segmental duplications and known copy number variants (SI Appendix), as well as regions under recent positive selection (49). Among contiguous genomic regions of at least 100 kb that satisfy these criteria, we then ranked targets for sequencing by their content of conserved and repetitive elements, and removed CpG islands. The NR dataset spans a total of 216,240 bp across 15 regions, each between 5,340 bp to 20,000 bp long (SI Appendix, Table S2). These and additional criteria for selecting regions are implemented in the Neutral Regions Explorer webserver at

http://nre.cb.bscb.cornell.edu (50).

**Sequencing, mapping and variant calling**

Illumina HiSeq 2000 with 100-bp paired-end reads was used for sequencing. Reads were mapped to hg18 human reference genome using BWA (51) (SI Appendix). For each individual, aligned reads were subjected to "duplicate removal" using Picard v.1.66 (http://picard.sourceforge.net). Subsequent SNV calling, quality control filtering, and genotype calling were performed with the Genome Analysis ToolKit, GATK-1.5-31 (36, 37), as detailed in SI Appendix. Analyses are based on 493 individuals that were successfully sequenced.

**Demographic inference**

To obtain estimates of recent demographic history, we calculated the likelihood of the observed site frequency spectrum (SFS) conditioned on several demographic models. To reduce parameter space, we fixed ancient history as estimated by previous studies (9, 14, 15) and only estimated parameters of more recent history. These models of recent history have either a single epoch of recent growth or an earlier epoch of growth preceding the recent growth. Models include different combinations of the following parameters: the time recent growth started, final $N_e$ after recent growth, $N_e$ before growth, start time of earlier growth, and $N_e$ after earlier growth (Table 1). We tested whether a model provides a better fit than another using Vuong's test (52). For each model, we estimated the SFS at different grid points using ms (53), with each grid point being a particular combination of parameter values. We then calculated the composite likelihood of the model following the approach of (9), as the probability of the observed minor allele (the less common of the two alleles) counts of all SNVs while accounting for missing data (SI Appendix). We profiled the likelihood surface using for each parameter 7-16 predefined grid points that span a range of plausible values. We increased the number of grid points for each parameter 10-fold by fitting a smooth spline function for the proportion of SNVs of each allele count as a function of all parameters, which improved accuracy (SI Appendix, Fig. S10), with only a minor increase in computational burden. Two-sided 95% confidence interval (CI) for each parameter was estimated following a $\chi^2_{(1)}$ distribution that accounts for variation across SNVs (SI Appendix).

# Acknowledgements

We thank Srikanth Gottipati and Aaron Sams for helpful advice, Matt Rasmussen for code to parse Newick trees (SPIMAP), and the editor and reviewers of this manuscript for helpful suggestions. This work was supported in part by NIH grants GM065509, HG003229 and HG005715. E.G was supported in part by a Cornell Center for Comparative and Population Genomics fellowship. A.K. was also supported by The Ellison Medical Foundation, an Alfred P. Sloan Research Fellowship, and the Edward Mallinckrodt, Jr. Foundation.

# REFERENCES

1. Cohen JE (1996) *How Many People Can the Earth Support?* (W. W. Norton & Company, New York) 1 Ed.
2. Roberts L (2011) 9 billion? *Science* 333(6042):540-543.
3. Haub C (1995) How many people have ever lived on earth? *Popul Today* 23(2):4-5.
4. Keinan A & Clark AG (2012) Recent explosive human population growth has resulted in an excess of rare genetic variants. *Science* 336(6082):740-743.
5. Hartl D & Clark A (2007) *Principles of population genetics* (Sinauer, Sunderland, MA).
6. Erlich HA, Bergstrom TF, Stoneking M, & Gyllensten U (1996) HLA Sequence Polymorphism and the Origin of Humans. *Science* 274(5292):1552b-1554b.
7. Garrigan D & Hammer MF (2006) Reconstructing human origins in the genomic era. *Nat Rev Genet* 7(9):669-680.
8. Harding RM, *et al.* (1997) Archaic African and Asian lineages in the genetic ancestry of modern humans. *Am J Hum Genet* 60(4):772-789.
9. Keinan A, Mullikin JC, Patterson N, & Reich D (2007) Measurement of the human allele frequency spectrum demonstrates greater genetic drift in East Asians than in Europeans. *Nat Genet* 39(10):1251-1255.
10. Takahata N (1993) Allelic genealogy and human evolution. *Mol Biol Evol* 10(1):2-22.
11. Yu N, *et al.* (2001) Global patterns of human DNA sequence variation in a 10-kb region on chromosome 1. *Mol Biol Evol* 18(2):214-222.
12. Tenesa A, *et al.* (2007) Recent human effective population size estimated from linkage disequilibrium. *Genome Res* 17(4):520-526.
13. Gutenkunst RN, Hernandez RD, Williamson SH, & Bustamante CD (2009) Inferring the joint demographic history of multiple populations from multidimensional SNP frequency data. *PLoS Genet* 5(10):e1000695.
14. Gravel S, *et al.* (2011) Demographic history and rare allele sharing among human populations. *Proc Natl Acad Sci U S A* 108(29):11983-11988.
15. Schaffner SF, *et al.* (2005) Calibrating a coalescent simulation of human genome sequence variation. *Genome Res* 15(11):1576-1583.
16. Mele M, *et al.* (2012) Recombination gives a new insight in the effective population size and the history of the old world human populations. *Mol Biol Evol* 29(1):25-30.
17. Nelson MR, *et al.* (2012) An abundance of rare functional variants in 202 drug target genes sequenced in 14,002 people. *Science* 337(6090):100-104.
18. Coventry A, *et al.* (2010) Deep resequencing reveals excess rare recent variants consistent with explosive population growth. *Nat Commun* 1:131.
19. Tennessen JA, *et al.* (2012) Evolution and functional impact of rare coding variation from deep sequencing of human exomes. *Science* 337(6090):64-69.
20. Maruyama T (1974) The age of a rare mutant gene in a large population. *Am J Hum Genet* 26(6):669-673.
21. Maruyama T (1974) The age of an allele in a finite population. *Genetical research* 23(2):137-143.
22. Kiezun A, *et al.* (2013) Deleterious alleles in the human genome are on average younger than neutral alleles of the same frequency. *PLoS Genet* 9(2):e1003301.

23. Kiezun A, *et al.* (2012) Exome sequencing and the genetic basis of complex traits. *Nat Genet* 44(6):623-630.
24. Gazave E, Chang D, Clark AG, & Keinan A (2013) Population Growth Inflates the Per-Individual Number of Deleterious Mutations and Reduces Their Mean Effect. *Genetics*.
25. Chamary JV & Hurst LD (2005) Evidence for selection on synonymous mutations affecting stability of mRNA secondary structure in mammals. *Genome biology* 6(9):R75.
26. Chamary JV, Parmley JL, & Hurst LD (2006) Hearing silence: non-neutral evolution at synonymous sites in mammals. *Nat Rev Genet* 7(2):98-108.
27. Fu W, *et al.* (2013) Analysis of 6,515 exomes reveals the recent origin of most human protein-coding variants. *Nature* 493(7431):216-220.
28. Barreiro LB, Laval G, Quach H, Patin E, & Quintana-Murci L (2008) Natural selection has driven population differentiation in modern humans. *Nat Genet* 40(3):340-345.
29. Waldman YY, Tuller T, Keinan A, & Ruppin E (2011) Selection for translation efficiency on synonymous polymorphisms in recent human evolution. *Genome biology and evolution* 3:749-761.
30. Novembre J, *et al.* (2008) Genes mirror geography within Europe. *Nature* 456(7218):98-101.
31. Humphreys K, *et al.* (2011) The genetic structure of the Swedish population. *PLoS One* 6(8):e22547.
32. Price AL, *et al.* (2009) The impact of divergence time on the nature of population structure: an example from Iceland. *PLoS Genet* 5(6):e1000505.
33. Di Gaetano C, *et al.* (2012) An overview of the genetic structure within the Italian population from genome-wide data. *PLoS One* 7(9):e43759.
34. The ARIC Investigators (1989) The Atherosclerosis Risk in Communities (ARIC) Study: design and objectives. The ARIC investigators. *Am J Epidemiol* 129(4):687-702.
35. Nelson MR, *et al.* (2008) The Population Reference Sample, POPRES: a resource for population, disease, and pharmacological genetics research. *Am J Hum Genet* 83(3):347-358.
36. McKenna A, *et al.* (2010) The Genome Analysis Toolkit: a MapReduce framework for analyzing next-generation DNA sequencing data. *Genome Res* 20(9):1297-1303.
37. DePristo MA, *et al.* (2011) A framework for variation discovery and genotyping using next-generation DNA sequencing data. *Nat Genet* 43(5):491-498.
38. Morrison AC, *et al.* (2013) Whole-genome sequence-based analysis of high-density lipoprotein cholesterol. *Nat Genet* 45(8):899-901.
39. NHLBI GO Exome Sequencing Project (ESP) Exome Variant Server. 2012(December).
40. Cooper GM, *et al.* (2005) Distribution and intensity of constraint in mammalian genomic sequence. *Genome Res* 15(7):901-913.
41. Myers S, Fefferman C, & Patterson N (2008) Can one learn history from the allelic spectrum? *Theor Popul Biol* 73(3):342-348.
42. Betzig L (2012) Means, variances, and ranges in reproductive success: comparative evidence. *Evolution and Human Behavior* 33(4):309-317.
43. Boyko AR, *et al.* (2008) Assessing the evolutionary impact of amino acid mutations in the human genome. *PLoS Genet* 4(5):e1000083.
44. Yu F, *et al.* (2009) Detecting natural selection by empirical comparison to random regions of the genome. *Hum Mol Genet* 18(24):4853-4867.

45. Ayodo G, *et al.* (2007) Combining evidence of natural selection with association analysis increases power to detect malaria-resistance variants. *Am J Hum Genet* 81(2):234-242.
46. Altshuler DM, *et al.* (2010) Integrating common and rare genetic variation in diverse human populations. *Nature* 467(7311):52-58.
47. Abecasis GR, *et al.* (2012) An integrated map of genetic variation from 1,092 human genomes. *Nature* 491(7422):56-65.
48. Price AL, *et al.* (2006) Principal components analysis corrects for stratification in genome-wide association studies. *Nat Genet* 38(8):904-909.
49. Akey JM (2009) Constructing genomic maps of positive selection in humans: where do we go from here? *Genome Res* 19(5):711-722.
50. Arbiza L, Zhong E, & Keinan A (2012) NRE: a tool for exploring neutral loci in the human genome. *Bmc Bioinformatics* 13.
51. Li H & Durbin R (2009) Fast and accurate short read alignment with Burrows-Wheeler transform. *Bioinformatics* 25(14):1754-1760.
52. Vuong QH (1989) Likelihood Ratio Tests for Model Selection and Non-Nested Hypotheses. *Econometrica* 57(2):307-333.
53. Hudson RR (2002) Generating samples under a Wright-Fisher neutral model of genetic variation. *Bioinformatics* 18(2):337-338.

# Figure Legend

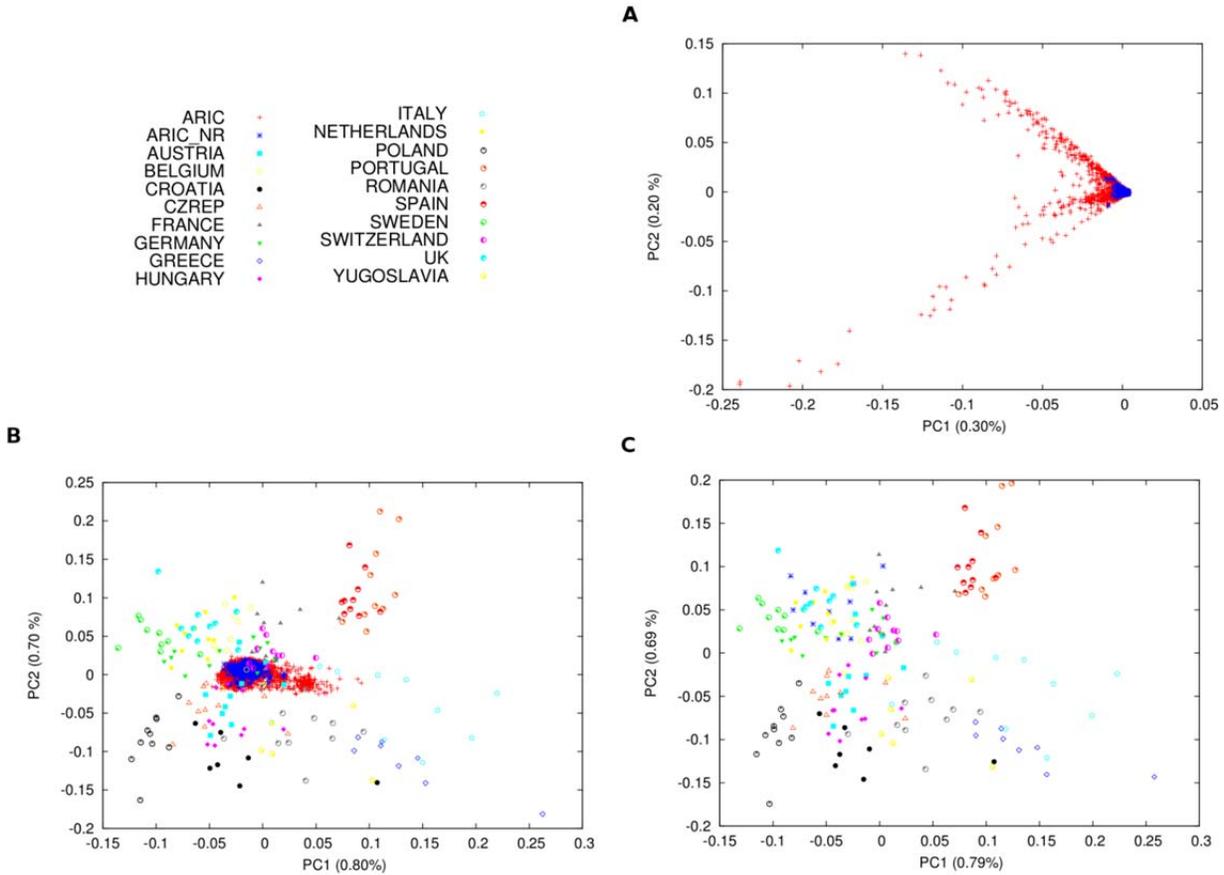

**Figure 1. Genetic ancestry of the NR sequenced individuals.** (**A**) Principal component analysis (PCA) of all individuals of European ancestry from the ARIC cohort, with the exception of outlier individuals (SI Appendix). Principal components 1 and 2 are plotted and reveal extensive population structure. Individuals chosen for sequencing in the present study are denoted in blue. (**B**) PCA of individuals from the POPRES cohort, with the individuals from panel **A** subsequently projected onto the resulting principal components (SI Appendix). Principal component 1 (*x-axis*) appears to capture southern *vs.* northern European ancestry, while principal component 2 (*y-axis*) western *vs.* eastern European ancestry, in line with the results of Novembre *et al.* (30). The full set of individuals from ARIC ('ARIC') show greater variability, mostly across the first principal component, than the set of individuals sequenced in this study ('ARIC_NR'). (**C**) Same as panel **B**, except that the PCA includes 10 randomly chosen individuals from the present study ('ARIC_NR'), rather than projecting all of them as in panel **B**. These 10 individuals cluster with individuals from the POPRES UK and related populations, which is also the case for other random sets of 10 ARIC_NR individuals.

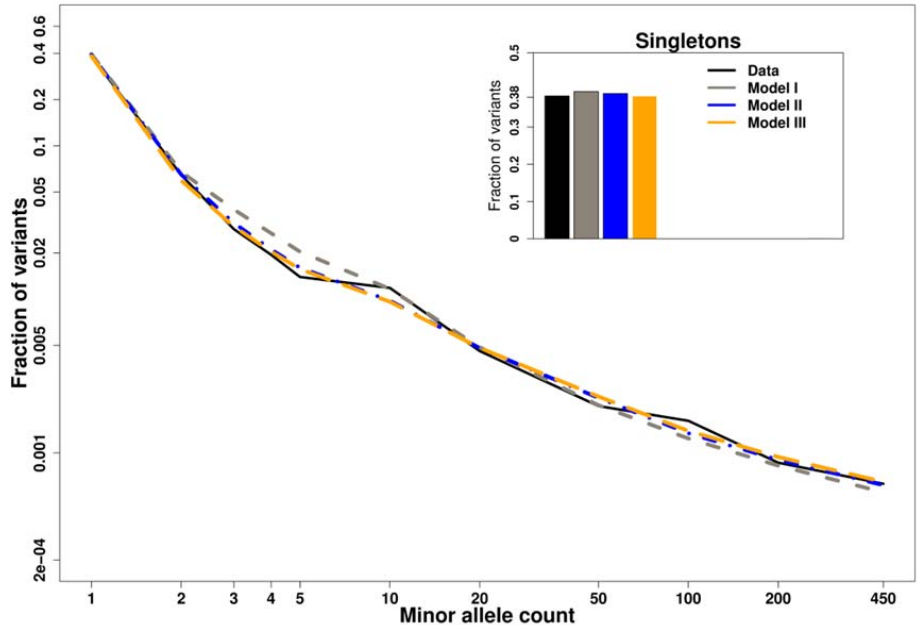

**Figure 2. SFS of the NR dataset and demographic models.** *X-axis* represents, in log scale, a partition of the number of copies of the minor allele with each number indicating the upper bound of a bin. (Minor allele counts of 1 through 5 are not binned.) *Y-axis* presents, in log scale, the proportion of SNVs that fall into each bin. Inset zooms in on the fraction of variants that are singletons (*y-axis* is presented in linear scale). 'Data' denotes the empirical SFS of NR data, 'Model I' a two-parameter model with one epoch of growth, where the duration of growth and final $N_e$ were estimated, 'Model II' an extension of Model I with a third parameter corresponding to $N_e$ before the growth epoch, and 'Model III' a model with two separate epochs of growth (Table 1). For clarity, Model IV is not presented since its SFS is very similar to that of Model III.

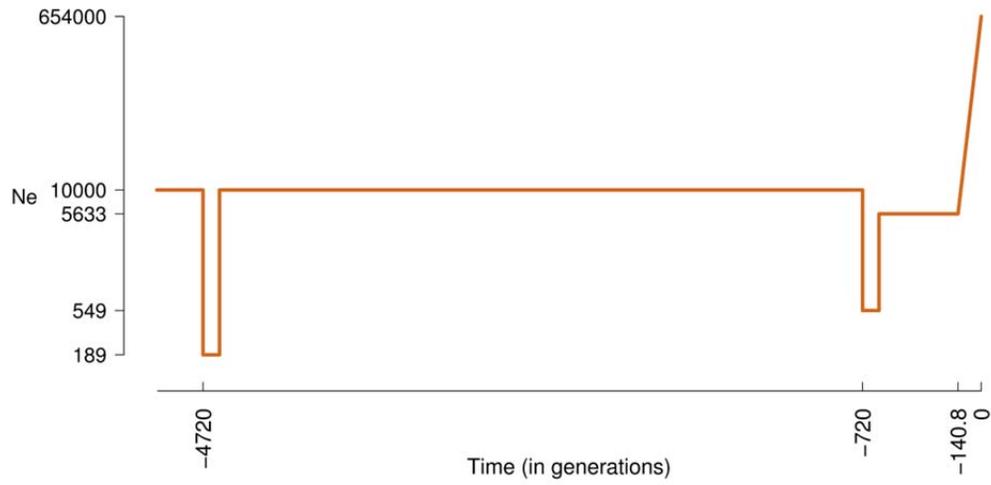

**Figure 3. Schematic representation of the best fit model.** Ne is shown in log scale during the last 5000 generations, with the last 620 generations as estimated by Model II (Table 1), and the preceding period following (9).

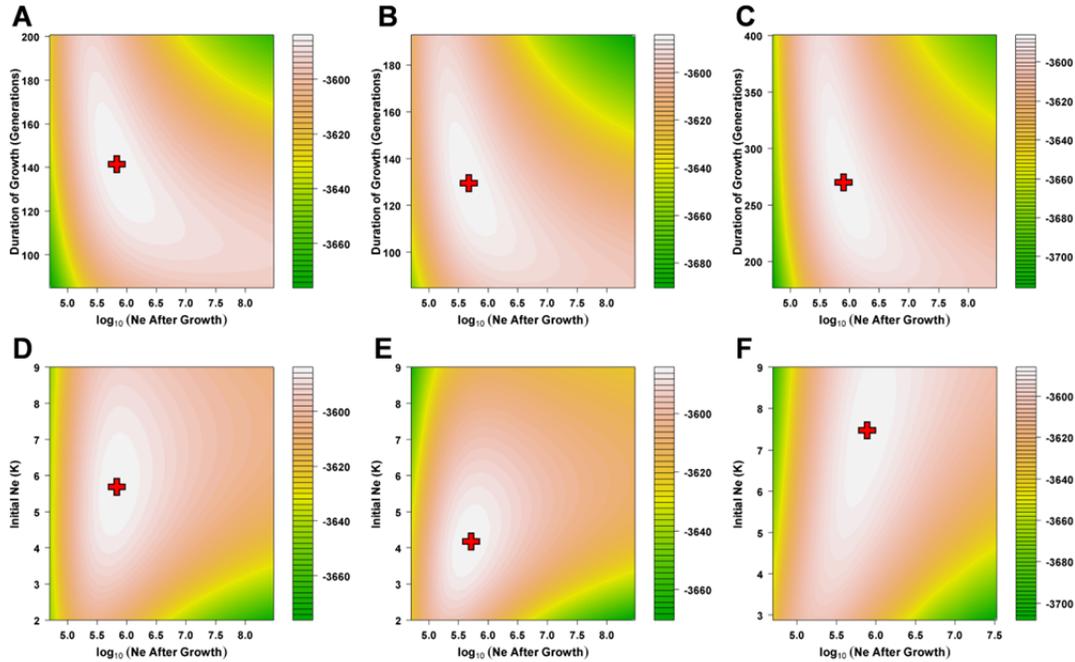

**Figure 4. Log-likelihood surface of Model II using three different models of ancient history.**
**(A-C)** Log-likelihood as a function of two parameters—the time growth started and the final $N_e$—with the third parameter ($N_e$ before the growth) fixed on the maximum likelihood estimate. **(D-F)** Similarly, with $N_e$ before the growth and final $N_e$ presented, and the time of growth fixed. The model was estimated by fitting these three parameters concurrently (Materials and Methods). **A** and **D** are for the model based on ancient demography from Keinan *et al.* (9) (Table 1), **B** and **E** from Gravel *et al.* (14), and **C** and **F** from Schaffner *et al.* (15) (SI Appendix, Table S5). Colored contours are at intervals of 2 log-likelihood. (Note the different scales between panels.) Red crosses denote the maximum-likelihood estimates.

**Table 1: Four models of recent demographic history and population growth.**

| Model* | Number of free parameters | Ne prior to growth | Duration of earlier growth (generations) | Ne after earlier growth | Duration of recent growth (generations) | Ne after recent growth (millions) | Growth rate during recent growth | Log likelihood |
|---|---|---|---|---|---|---|---|---|
| Model I | 2 | *10000* | N/A | N/A | **112.8 (92.9,136.8)** | **5.2 (0.8,300)** | 5.54% (3.2,11.1) | -3595.141 |
| Model II** | 3 | **5633 (4400,7100)** | N/A | N/A | **140.8 (116.8,164.7)** | **0.654 (0.3,2.87)** | 3.38% (2.4,5.1) | -3583.975 |
| Model III | 3 | *5633* | 267.3 | **5362 (3614,7955)** | **132.7 (101.8,165.5)** | **0.73 (0.3,5.7)** | 3.70% (2.6,6.4) | -3584.178 |
| Model IV | 4 | *5633* | **200 (200,600)** | **5000 (3000,15000)** | **140 (80,160)** | **0.5 (0.3, 50)** | 3.84% (0.36,4.44) | -3583.501 |

* The table describes the recent history estimated by four models, as well as the goodness of fit of each (log likelihood). **Bolded** values denote the maximum likelihood estimates (and 95% confidence intervals) of the free parameters estimated by each model. *Italicized* are values assumed as fixed in the model, and regular font denotes parameters that are a direct function of estimated parameters. All four models assume the model of ancient demographic history as estimated in (9), which includes two population bottlenecks (Fig. 3). Results based on other models of ancient demographic history are provided in SI Appendix.

** This model is considered as the best-fit model since none of the models with additional parameters provide a significantly improved fit.